\documentclass[aps,prl,floatfix,epsfig,twocolumn,showpacs,preprintnumbers]{revtex4}
\usepackage{amssymb}
\usepackage{graphicx}
\usepackage{amsmath}

\begin{document}

\title{High Surface Conductivity of Fermi Arc Electrons in Weyl semimetals }
\author{Giacomo Resta$^{\dag }$, Shu-Ting Pi$^{\dag }$, Xiangang Wan$^{\ast
} $, Sergey Y. Savrasov$^{\dag }$}
\affiliation{$^{\dag }$Department of Physics, University of California, Davis,
California, USA}
\affiliation{$^{\ast }$National Laboratory of Solid State Microstructures and Department
of Physics, Nanjing University, Nanjing, China}

\begin{abstract}
Weyl semimetals (WSMs), a new type of topological condensed matter, are
currently attracting great interest due to their unusual electronic states
and intriguing transport properties such as chiral anomaly induced negative
magnetoresistance, a semi--quantized anomalous Hall effect and the debated
chiral magnetic effect. These systems are close cousins of topological
insulators (TIs) which are known for their disorder tolerant surface states.
Similarly, WSMs exhibit unique topologically protected Fermi arcs surface
states. Here we analyze electron--phonon scattering, a primary source of
resistivity in metals at finite temperatures, as a function of the shape of
the Fermi arc where we find that the impact on surface transport is
significantly dependent on the arc curvature and disappears in the limit of
a straight arc. Next, we discuss the effect of strong surface disorder on
the resistivity by numerically simulating a tight binding model with the
presence of quenched surface vacancies using the Coherent Potential
Approximation (CPA) and Kubo--Greenwood formalism. We find that the limit of
a straight arc geometry is remarkably disorder tolerant, producing surface
conductivity that is a factor of 50 larger of a comparable set up with
surface states of TI. Finally, a simulation of the effects of surface
vacancies on TaAs is presented, illustrating the disorder tolerance of the
topological surface states in a recently discovered WSM material.
\end{abstract}

\maketitle

\section{I. Introduction}

While Weyl semimetals have long been studied for their condensed matter
realization of relativistic massless (Weyl) fermions\cite{Volovik}, and
related particle physics phenomenon such as the chiral anomaly\cite{Nielsen}%
, only recently have they attracted great interest\cite%
{Wan,Yang,Burkov,XiDaiTaAs,HasanTaAs,HosurReview,AshvinReview} for extending
concepts of Topological Insulators \cite{HasanRMP, MooreNature,PhysicsToday}
(TI) to gapless systems.

\begin{figure}[tbp]
\includegraphics[width=0.4\textwidth]{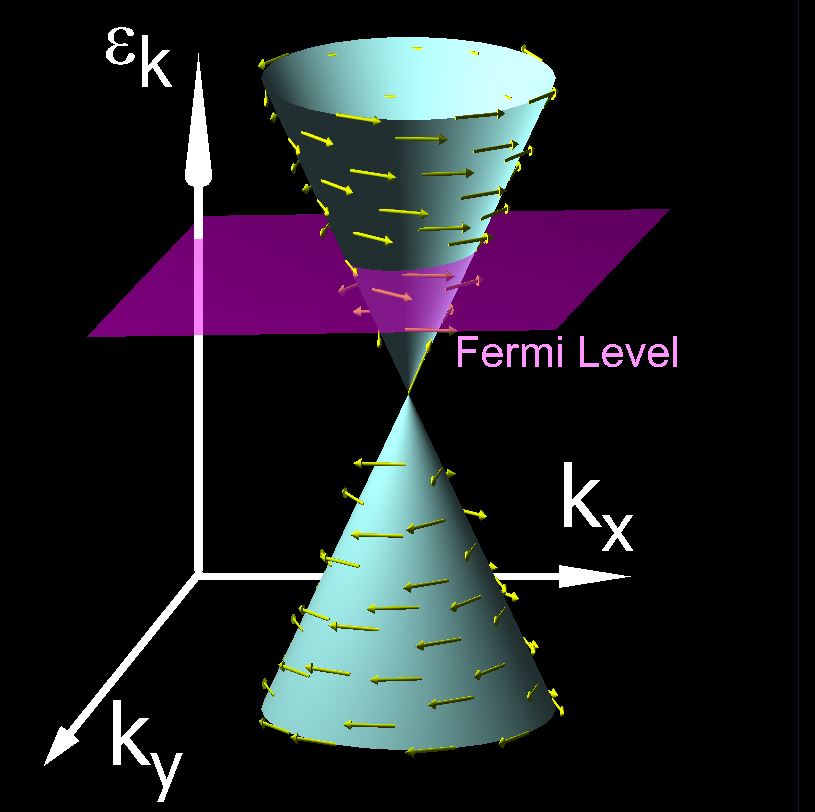} %
\includegraphics[width=0.4\textwidth]{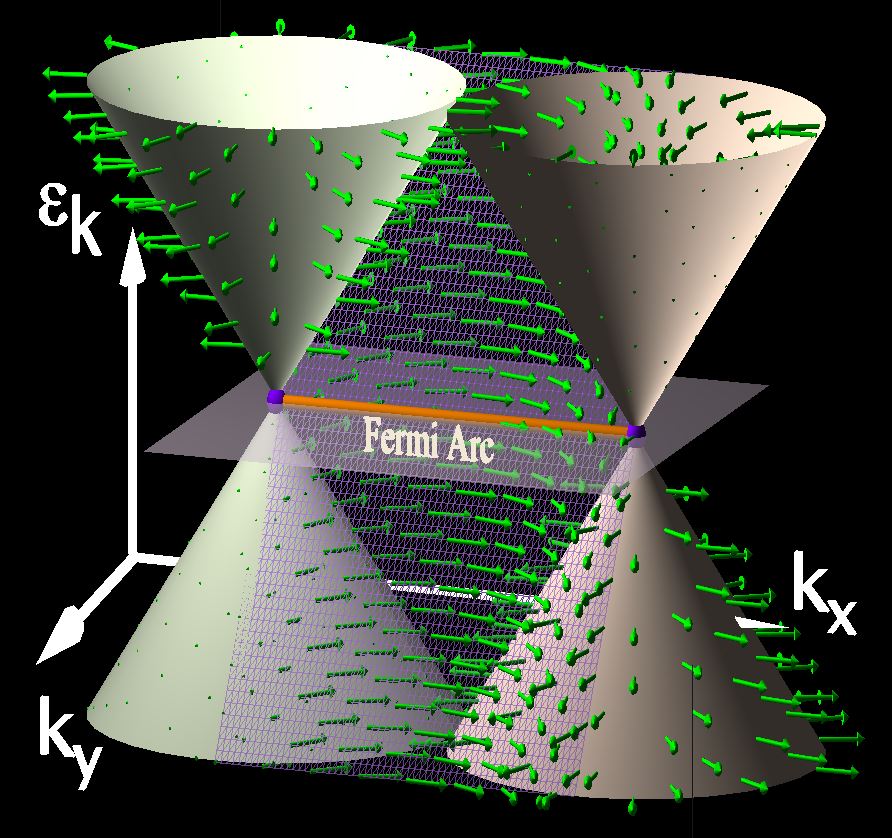}
\caption{(a) Dirac cone surface state of an idealized topological insulator
where the spin of the particle is tangential to the momentum, a phenomenon
known as spin--momentum locking. (b) The bulk and surface states for a Weyl
semimetal. The bulk band structure is fully gapped except at a pair of Weyl
nodes of opposite chiralities distinguishing varying spin orientations. The
surface state spans between the bulk Weyl cones with its spin states veering
from one cone to another. With the Fermi energy pinned at the Weyl nodes,
the surface state produces a Fermi arc connecting two Weyl nodes. }
\label{Fig1}
\end{figure}

Topological Insulators (TIs) are time--reverse symmetric states of matter
characterized by an insulating bulk which exhibits unique highly conductive
metallic surface states. In the case of 3D TIs, the surface states consist
of a massless 2D Dirac fermion dispersion relation with characteristic
helical spin--momentum locking and a circular Fermi surface (Figure~\ref%
{Fig1}a). This spin--momentum locking results in the suppression of
backscattering (k to --k) from non--magnetic impurities since states with
opposite spin and momentum remain orthogonal as along as time--reverse
symmetry is respected. As a result, the surface transport of 3D TIs is
uniquely robust to non--magnetic disorder.

In contrast, Weyl semimetals can only occur if either time--reversal or
inversion symmetry is broken and are characterized by a bulk band structure
which is fully gapped except at isolated points where the electronic
structure is represented by the 2x2 Weyl Hamiltonian. Each Weyl point acts
as a positive or negative monopole of Berry flux depending on whether its
chirality is +1 or -1. Since the total Berry flux through the WSM Brillouin
Zone must be zero, Weyl points always appear in pairs of opposite chirality%
\cite{NNTheorem}. The surface electronic states of WSMs are subsequently
described by arc states (Figure \ref{Fig1}b) which are topologically
protected as long as the bulk 3D Weyl points remain intact\cite{Wan}.

The topological protection in WSMs means that the Weyl points cannot be
gapped by infinitesimal perturbations and can only disappear by merging with
a Weyl point of opposite chirality in which case they form a double
degenerate Dirac point\cite{DSM} that is generally susceptible to small
perturbations. Depending on the path the Weyl points take in reciprocal
space, the arc states can either collapse to a point, or form a closed
contour, where gapping the resulting Dirac points may result in either a
trivial or topological insulator phase.

There are always two arcs that connect each set of Weyl points due to the
existence of two (top and bottom) surfaces of the material. They can, for
example, manifest themselves in quantum oscillatory techniques\cite%
{QuantumOscillations}. While there is no general spin--momentum locking
mechanism for the Fermi arc states, backscattering of electrons from the arc
associated with the top surface to the arc associated with the bottom
surface is forbidden due to zero spatial overlap between the corresponding
electronic states. Hence, considering transport perpendicular to the
orientation of the arcs, the situation is similar to the case of TIs.

The unique electronic states in Weyl semimetals, along with their
topological origin, makes it appealing to study their charge transport
mechanism. This was the subject of intense study lately\cite%
{Burkov,AshvinTransport,Balatsky,DiffusiveTransport,QuantumTransport,ALTransport,CriticalTransport,Carbotte}
where Weyl points have been found to be robust against weak bulk disorder
although it was argued that going beyond perturbative RG approaches may
induce a small density of states even at weak disorder\cite%
{Nandkishore,Pixley}. With the bulk density of states disappearing as $%
g(\epsilon )\sim \epsilon ^{2}/v_{F}^{3}$ where $\epsilon $ is the energy of
the Weyl points away from the nodal point and $v_{F}$ is the velocity of the
Fermi electrons, the bulk DC conductivity $\sigma $ in a clean free fermion
limit is expected to be zero\cite{Wan} at zero temperature $T$ and $\epsilon
=0$, although disorder, interactions, and thermal excitations modify this
result. With a characteristic strength of the impurity potential $V_{imp}$,
the scattering rate for the electrons $Im\Sigma (\epsilon )\sim
V_{imp}g(\epsilon )$ in Born approximation and the conductivity $\sigma \sim
v_{F}^{2}g(\epsilon )/Im\Sigma (\epsilon )$ remains finite even at $%
\epsilon =0$\cite{Burkov}. Interactions alone were found $\sigma $ to
acquire a linear temperature dependence\cite{AshvinTransport} and reproduce
a clean fermion limit$.$ A more complete study of bulk dc transport and ac
optical properties of 3D Dirac and Weyl semimetals including their
temperature dependence and doping away from the nodal point was recently
elaborated in details\cite{Carbotte}.

The existence of the topologically protected surface states leads to another
set of interesting phenomena. For example, the surface conductivity in 3D
TIs was extensively investigated\cite%
{DasSarma,Ong,StrongDisorder,Bi2Se3,Sinova,Austin}, and its disorder
tolerance has been emphasized. The Fermi arcs in WSMs gives rise to a
non--zero anomalous Hall effect with a semiquantized value of the Hall
conductivity proportional to the distance between the Weyl points\cite%
{Yang,Burkov}. The impurity scattering via the Fermi arcs has been studied
in dilute bulk disorder limit using Born approximation in a most recent work%
\cite{ArcTransport} where the dissipative nature of the surface currents in
WSMs has been highlighted and the effect of surface--to--bulk scattering was
emphasized.

In order to examine the conductivity of the Fermi arc electrons in Weyl
semimetals, we need to introduce a notion of surface conductivity in
general. In a standard setup, a homogeneous electric field is applied across
the sample, and the conductivity relates the current density as a linear
response to the applied field. We now wish to find the current density that
appears as the response to the electric field existing in a part of space,
such as a sample surface. A general linear response relationship is given by%
\begin{equation*}
j_{\alpha }(\mathbf{r},\omega )=\sum_{\beta }\int_{surface}\sigma _{\alpha
\beta }^{(surf)}(\mathbf{r},\mathbf{r}^{\prime },\omega )E_{\beta }\left( 
\mathbf{r}^{\prime },\omega \right) d\mathbf{r}^{\prime }
\end{equation*}%
where we restrict the field $E_{\beta }\left( \mathbf{r},\omega \right) $ by
a part of space. A standard Kubo--Greenwood approach applies for very
general perturbations \cite{Kubo}. This results in the same technicalities
in computing $\sigma _{\alpha \beta }^{(surf)}$ as for the bulk conductivity
with the matrix elements of electron velocity operators now restricted by
the area where $E_{\beta }\left( \mathbf{r},\omega \right) \neq 0$. For an
insulator, such as the 3D TI, this brings nothing new, since the current in
the system with bulk energy gap can only be carried by the surface Fermi
electrons, and there is no need to restrict the applied field by the sample
surface. However, in a WSM, spatially homogeneous fields will always produce
a non--zero bulk current due to thermal excitations, disorder and electronic
correlations even in ideal scenarios with the Fermi level pinned at the Weyl
points\cite{Burkov,AshvinTransport}. In thermodynamic limit, such current
will overwhelm all surface effects. Physically, however, we expect that the
Fermi arc electronic states extending well into the bulk near the Weyl
points, are capable of supporting strong surface currents, therefore
understanding their scattering cross sections both within the surface and
into the bulk is an important problem that we address in the present work.

\begin{figure}[tbp]
\includegraphics[width=0.4\textwidth]{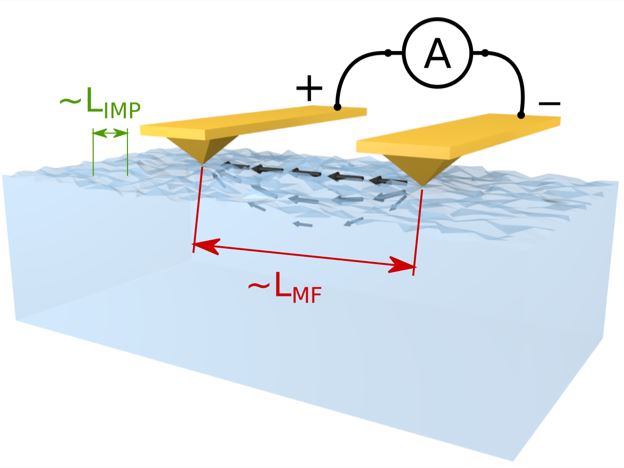}
\caption{Double--tip STM design\protect\cite{TwoTipsSTM} allows one to
measure surface charge transport at low temperatures. Two STM tips are
scanned sufficiently close to each other so that their distance is
comparable to the temperature dependent part of the mean free path $L_{MF}$,
but much larger than the distance between surface impurities $L_{imp}.$ A
sufficiently small voltage should be applied so that only a small layer near
the surface contributes to the current.}
\label{Fig9}
\end{figure}

An experimental setup that can probe surface conductivity can be easily
sketched. We imagine a double--tip Scanned Tunneling Microscope (STM) design%
\cite{TwoTipsSTM} as illustrated in Figure \ref{Fig9}. Two STM tips are
scanned across the surface of the WSM sufficiently close to each other so
that the temperature dependent part of the mean path $L_{MF}$ of the
electrons at the surface is comparable with the distance between the tips
but much larger than the average distance between surface impurities $%
L_{imp} $. This can always be achieved at sufficiently low temperatures. The
applied voltage between the tips has to be sufficiently small, so that the
corresponding electric field decaying quickly into the bulk will disturb
only a thin layer close to the surface. As two tips traverse through the
WSM\ surface, the local disorder affects the mean free path of the surface
states, and the robustness of the Fermi arc electrons can be directly probed.

Here, we focus on the surface charge transport in WSMs and discuss two major
relaxation mechanisms contributing to it. We first present a qualitative
discussion of electron--phonon scattering where we find that its
contribution strongly depends on the shape of the Fermi arc and becomes\
vanishingly small in the limit of a straight arc. This conclusion is
virtually independent of the particular form of the electron--phonon matrix
elements, but results from the fact that, in the limit of a straight arc,
scattering only occurs between states with the same velocity along the
direction of the current. Next, we perform numerical studies of the effect
of strong surface disorder on the surface conductivity in WSMs. Using a
tight--binding model in the slab geometry, we utilize a Coherent Potential
Approximation (CPA)\cite{CPAReview}, a self--consistent theory that has been
widely used for studies of substitutional alloys at arbitrary
concentrations, to investigate the effect of quenched surface vacancies on
the electronic spectral functions. We then calculate the slab conductivity
using the Kubo--Greenwood formalism. Our model takes into account the
scattering processes within the arc and from the arc to the bulk Weyl
points. We find that the WSM model with a straight Fermi arc is remarkably
disorder tolerant, producing a surface conductivity that is about 50 times
larger than the comparable set--up with circular surface states of the 3D TI
model. By computing the optical reflectivity, we also find\ that our WSM\
model with straight arc behaves as an ideal polarizer for incident light.
Finally, a discussion is given for the applicability of our results to real
WSMs discovered recently\cite{XiDaiTaAs,HasanTaAs}. We apply the CPA
technique to study the effect of surface vacancies in TaAs where we find
that the topological Fermi arc states display a robustness to disorder in a
real material setting.

\section{II. Electron--Phonon Scattering}

The electron--phonon contribution to the resistivity of metals has long been
understood both qualitatively\cite{Allen} and using first principle
electronic structure calculations based on the density functional linear
response approach\cite{SavrasovEPI}. In the linear temperature regime, the
electron--phonon resistivity can be written as $\rho _{e-ph}(T)\propto
\lambda _{tr}T,$ where effects due to the Fermi electrons and their
interactions with phonons is contained in the transport electron--phonon
coupling constant $\lambda _{tr},$ which is given by the following integral
over the Brillouin Zone\cite{Allen}, 
\begin{eqnarray}
\lambda _{tr} &=&\sum_{\mathbf{q}}\lambda _{tr}(\mathbf{q}),
\label{LambdaTr} \\
\lambda _{tr}(\mathbf{q}) &=&\frac{\sum_{\mathbf{k}}(v_{\mathbf{k}\alpha
}-v_{\mathbf{k}+\mathbf{q}\alpha })^{2}|V_{\mathbf{kk}+\mathbf{q}%
}^{e-ph}|^{2}\delta (\epsilon _{\mathbf{k}})\delta (\epsilon _{\mathbf{k}+%
\mathbf{q}})}{\sum_{\mathbf{k}}v_{\mathbf{k}\alpha }^{2}\delta (\epsilon _{%
\mathbf{k}})}.  \notag
\end{eqnarray}%
Here, $\epsilon _{\mathbf{k}}$ and $v_{\mathbf{k}\alpha }$ are the energies
and velocities of Fermi electrons, and $V_{\mathbf{kk}+\mathbf{q}}^{e-ph}$
is the matrix element of the potential induced by the displacement of atoms
from their equilibrium positions associated with a phonon of wavevector $%
\mathbf{q}$ (we omit summations over phonon branches for simplicity).

Although electron--phonon resistivity calculations can be carried out for
real materials using first principles methods \cite{SavrasovEPI}, to gain
physical insight into how the Fermi arc geometry influences the resistivity,
we first consider a minimal model for a single arc consisting of a segment
of a circle of radius $r_{F}$ that carries a fixed density of states $%
g_{arc}=s_{F}/(4\pi ^{2}v_{F})$ where $s_{F}=r_{F}\theta $ is the length of
the arc, $\theta $ is the angle the arc spans along the circle and $v_{F}$
is the velocity of the arc electrons which is assumed to be the same as the
Fermi velocities of the bulk Weyl points. The limit of a straight arc is
obtained by sending $r_{F}$ to infinity, $\theta $ to zero and keeping the
length $s_{F}$ fixed, while the circular Fermi surface characteristic of 3D
TIs is recovered when $r_{F}$ is equal to the Fermi wavevector $k_{F}$, $%
\theta =2\pi $, and $g_{TI}=k_{F}/(2\pi v_{F})$. We assume the arc to be
symmetric with respect to y axis, and consider the applied electric field
along y--axis. Note that although this model reproduces a smooth transition
from the circular surface states of TI to the case of a straight Fermi arc,
it should not be interpreted as modeling a literal transition between a WSM
and TI since in an actual material, merging the Weyl points may collapse the
arc.

\begin{figure}[tbp]
\includegraphics[width=0.4\textwidth]{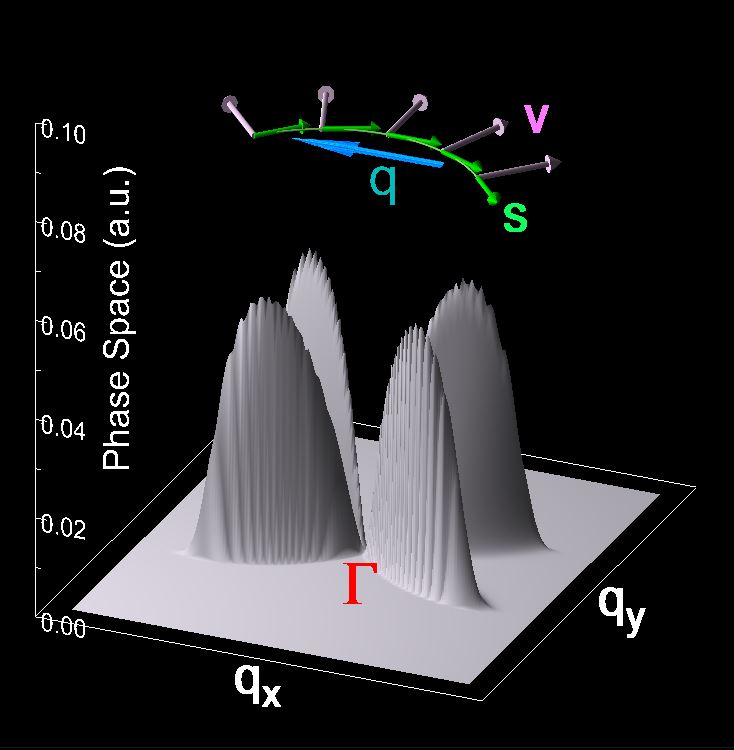} %
\includegraphics[width=0.4\textwidth]{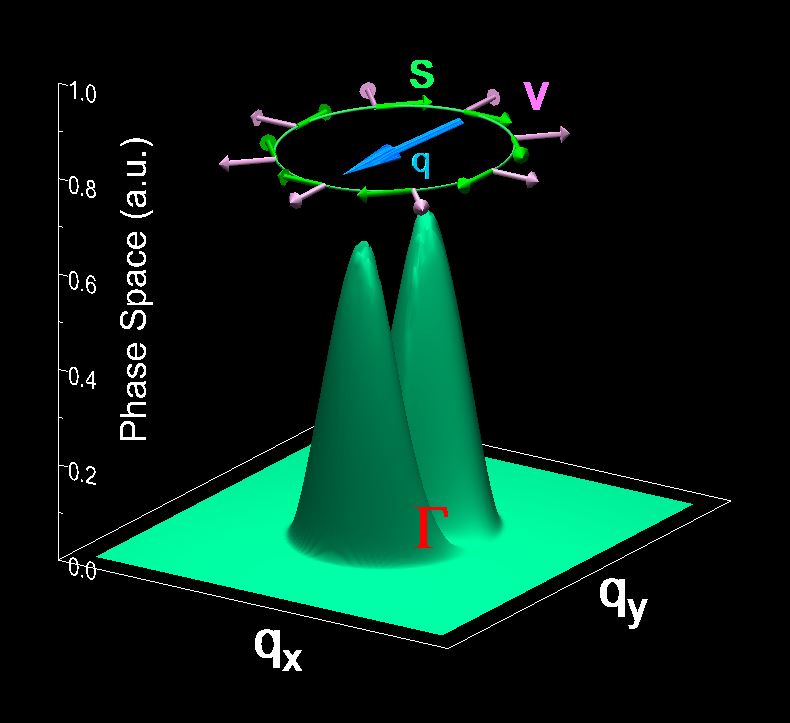}
\caption{Contributions to electron--phonon transport from the scattering
wavevector $\mathbf{q}$ for an arc with angles (a) $\protect\theta =\protect%
\pi /2$ and (b) $\protect\theta =2\protect\pi $ (a complete circle
representing the limit of 3D TI). Note difference in scales in two figures.
Note that contributions by 90 degree scattering processes are suppressed
(line corresponding to $q_{y}=0$) because initial and final states have the
same velocity along the direction of the current (y-axis). Also note that in
(b) 180 degree backscattering processes with scattering wavevector $q=(0,\pm
2k_{F})$, where $k_{F}=0.25$, are forbidden due to the orthogonality of
states with opposite spins.}
\label{Fig3}
\end{figure}

We can consider the scattering phase space which contributes to the
transport coefficient by taking the electron--phonon scattering matrix
element $V_{\mathbf{kk}+\mathbf{q}}^{e-ph}$ to be a constant in the
expression for $\lambda _{tr}(\mathbf{q}).$ However, to ensure that
spin--momentum locking is recovered in the TI\ limit where $\theta =2\pi $,
we assume that the spin is always tangential to the arc. For the isotropic
Weyl model illustrated in Figure \ref{Fig1}b this is approximately the case
near the middle of the arc. With this spin arrangement, the spin dependent
part of the matrix element $V_{\mathbf{kk}+\mathbf{q}}^{e-ph}$ is given by, 
\begin{equation}
V_{\mathbf{kk}+\mathbf{q}}^{spin}=\frac{1}{2}\left( \hat{k}\cdot (\widehat{%
k+q})+1\right) ,  \label{Vspin}
\end{equation}%
which reflects the overlap between spinor states at $\mathbf{k}$ and $%
\mathbf{k}+\mathbf{q}$. The $\mathbf{q}-$dependence of the phase space
integral $\sum_{\mathbf{k}}(v_{\mathbf{k}\alpha }-v_{\mathbf{k}+\mathbf{q}%
\alpha })^{2}|V_{\mathbf{kk}+\mathbf{q}}^{spin}|^{2}\delta (\epsilon _{%
\mathbf{k}})\delta (\epsilon _{\mathbf{k}+\mathbf{q}})$ can be evaluated
numerically. Figure \ref{Fig3} shows the result for an arc with $\theta =\pi
/2$ (Figure \ref{Fig3}a) and a complete circle with $\theta =2\pi $ (Figure %
\ref{Fig3}b). A remarkable observation is that in the limit of a straight
arc, the phase space collapses as contributions by 90 degree scattering
processes are all suppressed since electrons are scattered between states
with the same velocity along the direction of the current. This makes the
factor $v_{\mathbf{k}\alpha }-v_{\mathbf{k}+\mathbf{q}\alpha }=0$ in the
expression for $\lambda _{tr}(\mathbf{q})$ (where for the transport along
the y--axis we take $\alpha =y$). In the TI\ limit of the Fermi circle, one
sees that 180 degree backscattering processes with wavevector $\mathbf{q}%
=(0,\pm 2k_{F})$ are forbidden due to the orthogonality of spinor states
with opposite spin orientations and that contributions by 90 degree
scattering processes (described by $q_{y}=0$ in Fig \ref{Fig3}b) are
similarly suppressed since the velocity of the electrons projected onto the
direction of the current remains the same after scattering.

\begin{figure}[tbp]
\includegraphics[width=0.4\textwidth]{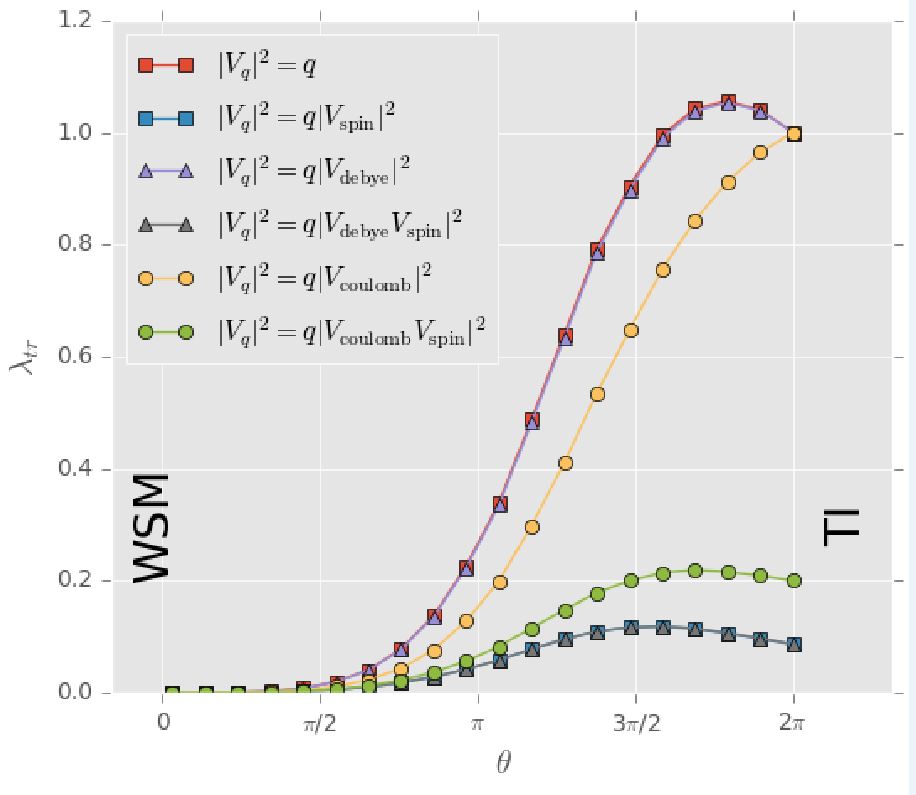}
\caption{Transport coupling constant $\protect\lambda _{tr}$ describing
intra--arc scattering processes as a function of the curvature of the Fermi
arc for various choices of the electron--phonon interaction matrix elements:
a constant, a Debye screened and bare Coulomb interactions as well as
incorporating the spin dependence of the arc states assuming an approximate
tangential arrangement. The normalization is chosen such that $\protect%
\lambda _{tr}=1$ for the complete circle representing the limit of 3D TI in
the absence of the spin structure.}
\label{Fig4}
\end{figure}

To provide a more realistic wave vector dependence for the electron--phonon
matrix elements we can consider a deformation potential type model with $|V_{%
\mathbf{kk}+\mathbf{q}}^{e-ph}|^{2}=D^{2}q$ and additionally include a Debye
model for screening in 2D%
\begin{equation}
V_{Debye}(\mathbf{q})\symbol{126}\frac{1}{q+\kappa _{D}},  \label{VDebye}
\end{equation}%
where the Debye screening constant is assumed to be $\kappa _{D}=2\pi
e^{2}g_{arc}$ (for simplicity we choose the case of a free relativistic
electron gas in 2D\cite{RelativisticGas2D} although one can also introduce
the static dielectric constant of the substrate material such as the bulk
TI). Figure \ref{Fig4} illustrates our calculated transport coupling
constant $\lambda _{tr}$ as a function of the arc curvature described by the
angle $\theta $ for several choices of the electron--phonon matrix elements $%
V_{\mathbf{kk}+\mathbf{q}}^{e-ph}$ which we represent it as a product of the
spin and spatial parts: $V_{\mathbf{kk}+\mathbf{q}}^{e-ph}=V_{\mathbf{kk}+%
\mathbf{q}}^{spin}V_{Space}(\mathbf{q}).$ We distinguish the case with $V_{%
\mathbf{kk}+\mathbf{q}}^{spin}=1$ and the case given by equation (\ref{Vspin}%
) to account for the spin structure of the electronic states along the arc.
We also distinguish cases with $|V_{Space}(\mathbf{q})|^{2}=q$, $|V_{Space}(%
\mathbf{q})|^{2}=q|V_{Debye}(\mathbf{q})|^{2}$ and $|V_{Space}(\mathbf{q}%
)|^{2}=q|V_{Bare}(\mathbf{q})|^{2}$ as the extreme limit of the 2D bare
Coulomb repulsion. The transport coupling constant is normalized to unity in
the limit of a complete circle ($\theta =2\pi $) when using $V_{\mathbf{kk}+%
\mathbf{q}}^{spin}=1.$

A few observations can be made. First, the transport coupling constant is
very small for arc geometries where $\theta <\pi /2$ and completely
disappears in the limit of a straight arc. Second, the spatial dependence of
the electron--phonon matrix elements as given by the constant, the Debye
screening and the bare Coulomb models, only weakly influences this result.
Third, the spin structure of the arc states leads to a reduction of the
scattering by a factor of 2--8.

We now discuss the low temperature dependence of the surface resistivity.
The 2D surface states of WSM should result in a $T^{4}$ behavior which
becomes linear in $T$ once it raises above the characteristic Debye energy
of phonons $\hbar \omega _{D}$. This is primarily due to the observation
that the energy of Fermi electrons in a metal is normally much larger than $%
\hbar \omega _{D}.$ In the opposite limit, the crossover occurs at the so
called Bloch--Gr\"{u}neisen temperature, which was, for example, a well
established case for graphene\cite{Efetov,Fuhrer} and could be realized for
the WSMs with anomalously small Fermi arcs. We provide complete analysis of
the temperature dependent electron--phonon scattering in Appendix A.

Our general conclusion is that the intra--arc contribution to the
electron--phonon resistivity in WSMs may become vanishingly small in the
limit that the arc approaches a straight line which makes this transport
mechanism more effective than the circular Fermi surfaces of the TIs.

\section{III. Impurity Scattering}

The contribution to the resistivity by electron--impurity scattering is
another important mechanism of charge transport in metals, and the influence
of bulk impurities on electronic scattering in Weyl systems has been
extensively studied in a number of recent works \cite%
{AshvinTransport,Balatsky,DiffusiveTransport,QuantumTransport,ALTransport,CriticalTransport,ArcTransport}%
. Since Weyl points remain intact in the presence of weak bulk disorder, an
important question arises on the influence of surface disorder, which is
known to have a significant impact on the surface transport properties of
such well--studied topological insulator system as Bi$_{2}$Se$_{3}$\cite%
{Bi2Se3Transport,Bi2Se3Transport2}.

In a real sample, surface disorder can be substantially strong, hence the
Born approximation which is typically used in the analysis of impurity
scattering may not be adequate. We therefore utilize the Coherent Potential
Approximation (CPA)\cite{CPAReview}, a self--consistent theory of alloys
that simulates random disorder up to arbitrarily strong concentrations. In
particular, the method allows us to simulate quenched surface vacancies
which is one of the primary forms of surface disorder in many materials. The
method explicitly incorporates scattering events between states within the
arc and from the arc to the bulk Weyl points which are slightly affected by
the presence of strong surface disorder. Note however that when the vacancy
concentration reaches 100\% (a complete removal of a single surface layer),
the electronic states should recover their values at zero concentration. One
of the major limitations of the CPA approach is the lack of Anderson
localization effects. Fortunately, this is not a serious issue when
considering massless fermions in 2D\cite{AL} although a strong bulk disorder
may lead to localization for 3D Dirac fermions\cite%
{ALTransport,CriticalTransport}.

For the WSM, we consider a realistic tight--binding model with only two bulk
Weyl points \cite{Turner}. Using a slab geometry that is infinite in xy
directions and finite in the z direction (c--axis) this model shows a single
straight Fermi arc that is extended between the Weyl points located at $%
k_{w}=(\pm 0.25,0)2\pi /a$ . To compare the result of our simulation with
the limit of a circular Fermi surface, we also consider a realistic
tight--binding model of TI \cite{TIShen} which shows a surface Dirac cone
located around the $\Gamma $ point. The parameters of both models are
adjusted so that the Fermi velocities and densities of states remain the
same in both simulations (See Appendix B for details). This setup allows us
to directly compare the effect of the arc geometry on the conductivities
which we determine using the Kubo--Greenwood formalism.

\begin{figure}[tbp]
\includegraphics[width=0.4\textwidth]{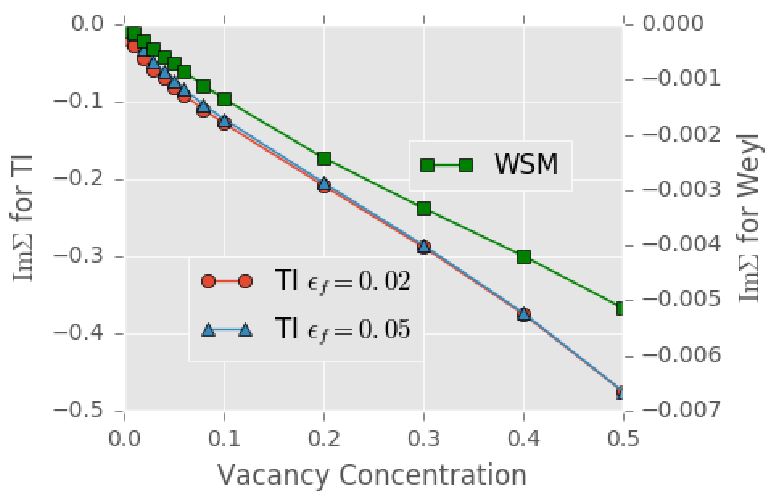}
\caption{The imaginary part of the self--energy corresponding to the surface
of the material for WSM and 3D TI as a function of the vacancy concentration
at the surface layer. The imaginary part of the self--energy for the WSM is
approximately two order of magnitude smaller than that for the 3D TI when
using models with the same Fermi velocities and densities of states. (The
position of the Fermi level for the TI\ should be at $\protect\epsilon %
_{f}=0.02$ to match the density of states of the WSM but we find a similar
result for a range of values up to $\protect\epsilon _{f}=0.05$)}
\label{Fig5}
\end{figure}

First, we discuss how the presence of surface disorder affects the spectral
functions. Figure\ \ref{Fig5} shows the imaginary part of the self--energy $%
Im\Sigma $ describing the life time effects of the surface states
that is obtained from our CPA\ simulation for both models as a function of
vacancy concentration $c$ on the top and bottom layers. We find that $Im\Sigma $ 
for the WSM model is approximately two orders of magnitude
smaller than the imaginary part of the self--energy for TI. This is partly
due to the fact that the TI Dirac cone is mainly a pure surface state\cite%
{Austin} while the arc states extend well into the bulk once they approach
Weyl points and become insensitive to the surface disorder. Our result
implies that the mean impurity scattering rate of the WSM surface is
significantly smaller than in TI. The Fermi level for the TI should be
placed at $0.02$ to match the density of states of the WSM. However, the
imaginary part of the self--energy is essentially independent of the Fermi
level in TI as long as we stay in the vicinity of the Dirac point (we show
two results for $\epsilon _{f}=0.02$ and $\epsilon _{f}=0.05$ on Figure \ref%
{Fig5}) as was discussed in a recent work\cite{Austin}.

\begin{figure}[tbp]
\includegraphics[width=0.4\textwidth]{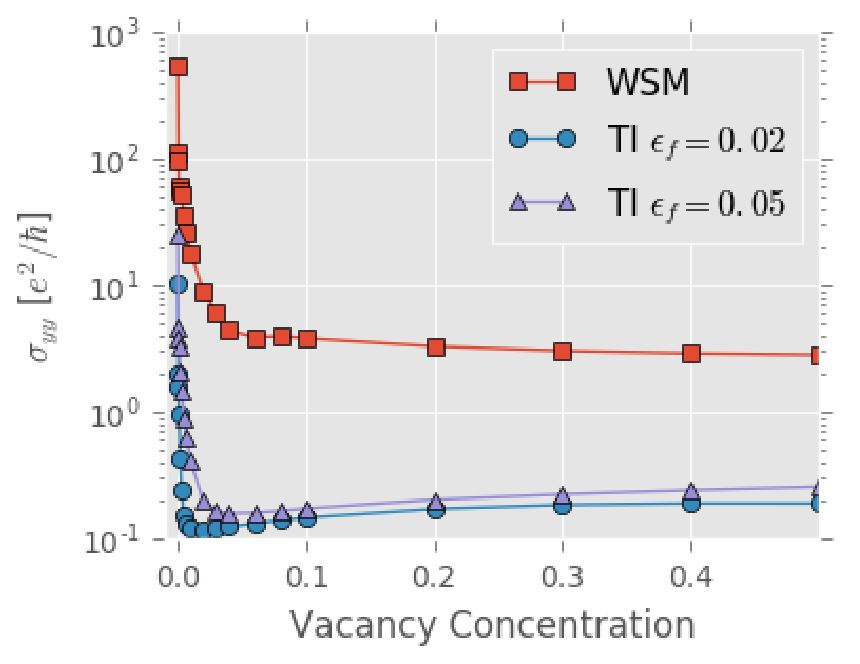}
\caption{Zero temperature DC conductivity as a function of surface vacancy
concentration along the axis perpendicular to the straight arc of the WSM
slab as well as a similar calculation for the TI slab for two values of the
Fermi level ($\protect\epsilon _{f}=0.02$ and $\protect\epsilon _{f}=0.05$).}
\label{Fig6}
\end{figure}

To provide a direct comparison between the charge transport properties in
the WSM and TI cases, we have used the Kubo--Greenwood formalism with the
self--energies obtained from our CPA\ simulation to calculate the slab
conductivities for both models. Figure~\ref{Fig6} gives the zero temperature
DC conductivity for the WSM as a function of surface vacancy concentration
for an electric field perpendicular to the orientation of the arc as well as
the conductivity for the TI. As a result of the smaller imaginary part of
the self--energy, and the suppression of scattering contributions for the
straight arc, the zero temperature DC conductivity of the WSM is much less
affected by surface vacancies than that for TI. We find the difference
between the two conductivities to be around 50. This result advances the
idea of realizing a more disorder tolerant surface charge transport than
that proposed for TIs as it appears that, in the limit of weak bulk
disorder, the Fermi arc surface states of WSMs are more robust to the
presence of surface vacancies than the surface states of TIs.

\begin{figure}[tbp]
\includegraphics[width=0.4\textwidth]{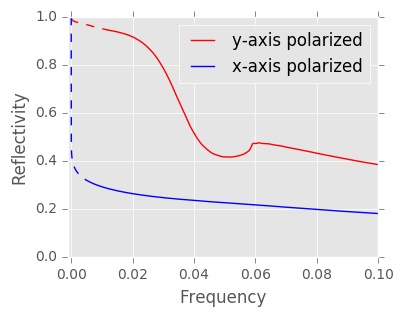}
\caption{Optical reflectivity for the WSM model with a surface vacancy
concentration of c = 0.1 for light which is linearly polarized along the
x-axis and y--axis. The Fermi arc state is parallel to the x-axis.}
\label{Fig7}
\end{figure}

The unique geometry of the straight arc with its Fermi velocities in the
perpendicular direction results in interesting consequences for the optical
properties of our WSM model. For the bulk Weyl and Dirac cone states, this
topic has been recently studied in detail both theoretically\cite{Carbotte}
as well as experimentally\cite{Chen} for a Dirac semimetal ZrTe$_{5}$.
However, accounting for the arc--to--bulk optical transitions alters the
response functions and results in strongly anisotropic optical properties.

Using the Kubo--Greenwood formalism, we have calculated the optical (AC)
conductivity for our WSM slab setup with a surface disorder of $c=0.1$ for
light polarized along the arc (x--axis) and in the perpendicular direction
(y--axis). Figure\ \ref{Fig7} plots the calculated reflectivity, where we
see that the reflectivity for y--polarized light exhibits a classical
metallic behavior while the reflectivity for x--polarized light is
semi--metallic, exactly as it was found in several recent works\cite%
{Carbotte,Chen}. Thus, a single straight arc acts as an ideal polarizer for
reflected light (for frequencies above phonons), an interesting effect with
possible applications in optics.

\section{IV. Results for Real Materials: TaAs Weyl Semimetal}

Physical insight gained from our simulations on models are now discussed in
the context of real materials that have been discovered in the past few
years to show WSM\ behavior. This broad class of solids includes systems
such as TaAs\cite{XiDaiTaAs,HasanTaAs,S1}, NbP\cite{S2}, TaP\cite{S3}, NbAs%
\cite{S4}, as well as so called type--II WSMs such as MoTe$_{2}$\cite{S5},
MoP$_{2}$, WP$_{2}$\cite{S6}, WTe$_{2}$\cite{XG1}, and LaAlGe\cite{S7}.
Despite sharing many common properties, such as a chiral anomaly induced
negative magnetoresistance effect\cite{S8}, their detailed topological
properties are very different. LaAlGe has 40 Weyl nodes and TaAs has 24
nodes while TaIrTe$_{4}$\cite{S9} shows only 4, the minimal number allowed
in an inversion symmetry breaking system. In addition, some WSMs such as TaAs%
\cite{S1} and NbAs\cite{S4} show very short and curved Fermi arcs, while
HgTe--class of WSM shows large circular Fermi arcs\cite{S10} and WSMs such
as Ta$_{3}$S$_{2}$\cite{S11}, TaIrTe$_{4}$\cite{S9} and Mo$_{x}$W$_{1-x}$Te$%
_{2}$ \cite{S12} show very long and straight Fermi arcs. A large diversity
here offers a unique platform to develop topological electronic devices\cite%
{S13}, and the guiding principles how the arc geometry affects its transport
properties can be very useful in real world scenarios.

Recently, several approaches have been proposed to engineer the shape and
length of the Fermi arcs by introducing appropriate doping\cite{S12}. While
bulk doping may lead to more significant contributions from arc--to--bulk
scattering and the multiplicity of arcs would necessarily assume inter--arc
scattering of the electrons, projecting the surface states onto certain
crystallographic directions or manipulating the arcs spin structures by
varying strength of spin--orbit coupling using substitutions\cite{S14} are
possible ways to suppress those effects. All this makes control of
topological surface transport a promising research direction.

To illustrate the robustness of the Fermi arcs to surface disorder in a real
material setting, we consider TaAs as an example. We perform ab initio
electronic structure of TaAs surface in a 6 unit--cell--long slab geometry
using Density Functional Theory in its Generalized Gradient Approximation.
We subsequently simulate the effect of quenched surface vacancies using
self--consistent CPA\ theory which we run for a range of concentrations to
explore the evolution of the electronic spectral functions (See Appendix C
for complete\ details). TaAs is a complex system and exhibits a variety of
states in the vicinity of the Fermi level including regular bulk Fermi
states, doped Weyl points as well as ordinary surface states and the Fermi
arcs. They all have been carefully mapped out by recent ARPES\cite%
{XiDaiTaAs,HasanTaAs}\ and quasiparticle interference\ \cite{S15}
experiments.

\begin{figure}[tbp]
\includegraphics[width=0.53\textwidth]{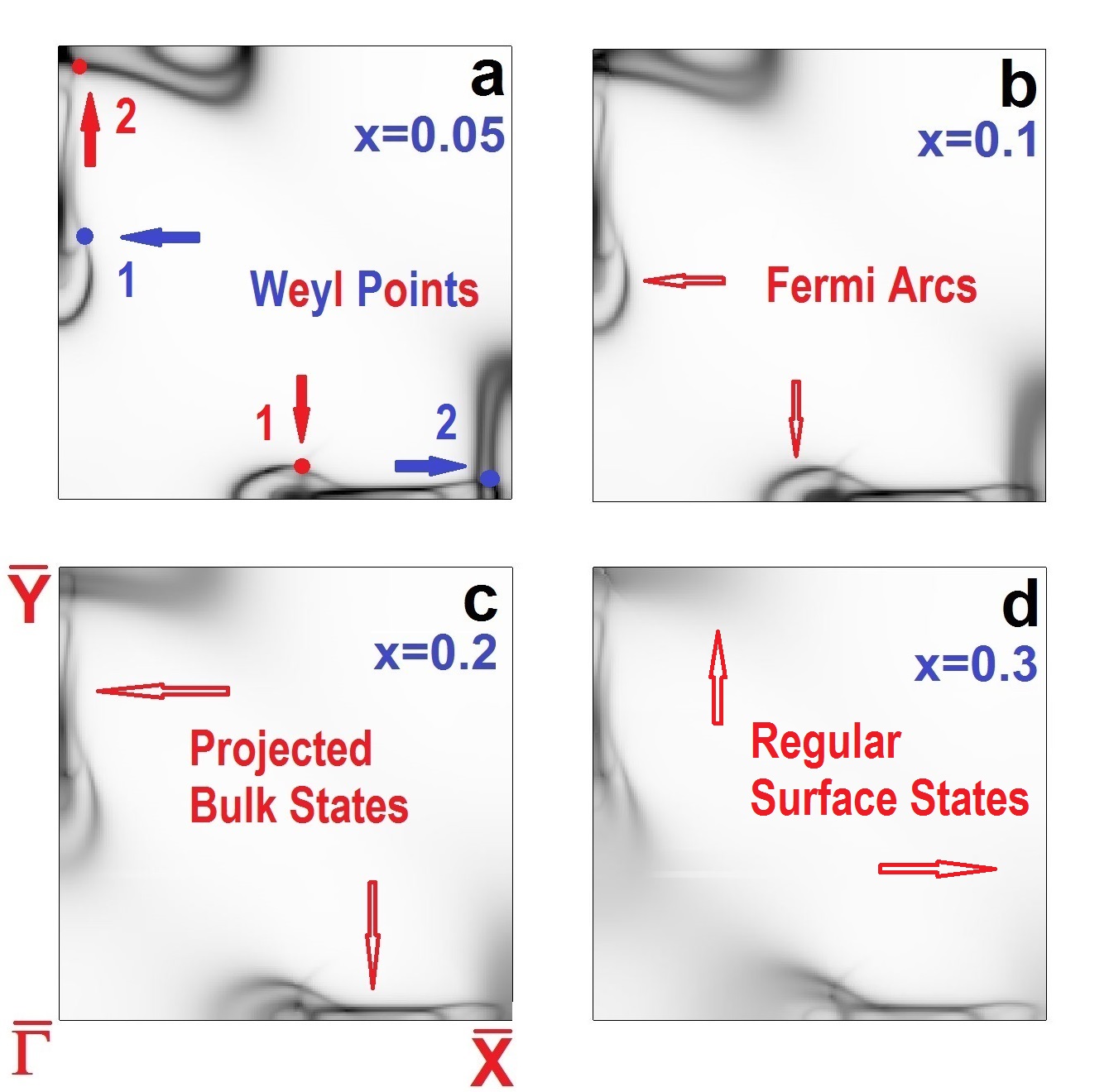}
\caption{The electronic states of the TaAs slab around the Fermi level in
the presence of x = 0.05 (a), x=0.1 (b), x=0.2 (c), and x=0.3 (d) of surface
vacancy concentrations at top As layer obtained using Coherent Potential
Approximation. We note that despite the presence of surface vacancies the
Fermi arcs are more disorder tolerant than regular surface states especially
in the vicinity of Weyl points. The regular bulk Fermi states projected on
the surface Brillouin Zone also remain largely unaffected by the surface
disorder.}
\label{Fig8}
\end{figure}

Figure \ref{Fig8} shows evolution of the Fermi states of the TaAs slab
structure that are projected onto the As terminated surface for various
concentrations of substitutional vacancies that we impose at topmost As
layer of the slab, ranging from $x$=0.05 to $x$=0.3, Figs.\ref{Fig8}(a--d).
\ For smaller concentrations of vacancies, x=0.05--0.1, \ref{Fig8}(a,b), we
note well defined (horseshoe--like) Fermi arcs connecting bulk Weyl points
that we call type 1. They have been widely discussed in recent TaAs
literature \cite{HasanTaAs,S15}. Also, there are other states (stretching
along $\bar{\Gamma}\bar{X}$ and $\bar{\Gamma}\bar{Y}$ lines) which are
composed from the bulk Fermi electrons that are projected onto the surface
Brillouin Zone, as well as bow--tie looking surface states seen at the
endings of those structures (in the vicinity of $\bar{X}$ and $\bar{Y}$
points). There should exist another set of Fermi arcs around $\bar{X}$ and $%
\bar{Y}$ points which originate from the Weyl points called type 2.
Unfortunately, there is some discrepancy in the current literature related
to the position of these arcs. Sometimes they were associated with straight
lines connecting the points 1 and 2, or with very short lines connecting the
only points 2 or with the bow--tie shaped structure around points $\bar{X}$
and $\bar{Y}$\cite{XiDaiTaAs,HasanTaAs,S15}. Nevertheless, it is clear that,
as the disorder increases, the surface states get broadened by the
self--energy effects which is seen for both regular surface states as well
as for the arc states. However, since the arc electrons are continuously
connected to the bulk Weyl points, the areas in the vicinity of the Weyl
points remains largely unaffected by disorder. This is in contrast to the
regular surface states which are expected to be more susceptible to
disorder. As there is no bulk disorder in our simulation, the bulk states
are largely unaffected by the surface vacancies.

\section{V. Conclusion}

In conclusion, we have analyzed the effects of electron--phonon scattering
and quenched surface vacancies on the surface charge transport properties
via the Fermi arcs. Our simulations on models showed that in the limit of a
straight arc and disorder free bulk, the contributions of both scattering
mechanisms to the resistivity are significantly suppressed and the Fermi
arcs can support a near dissipationless surface current. This allows us to
bring interesting parallels between 3D topological insulators and ideal Weyl
semimetals: The former are non conductors in the bulk and exhibit metallic
conductivities at the surface, while the latter show a bad metallic
conductivity at the bulk but high charge conductivity at the surface.

At the end, we also demonstrated that the Fermi arcs remain disorder
tolerant in TaAs. Although it is challenging to discuss the part of
conductivity connected to the Fermi arcs due to multiplicity of other
effects, such as thermal excitations, finite bulk disorder as well as
contributions from regular Fermi electrons, we suggest that either thin
films samples or experimental double--tip STM\ design can be useful for
studying surface charge transport mechanism in real Weyl semimetals.

\section{Acknowledgement}

The authors acknowledge useful conversations with Xi Dai, G. Kotliar, E. da
Silva Neto, W. Pickett, R. Singh, A. Vishwanath, H.M Weng, D. Yu. This work
was supported by the US National Science Foundation Grant DMR--1411336
(S.Y.S.).

\section{Appendix A. Semicircular Arc Model}

\begin{figure}[tbp]
\includegraphics[width=0.7\linewidth]{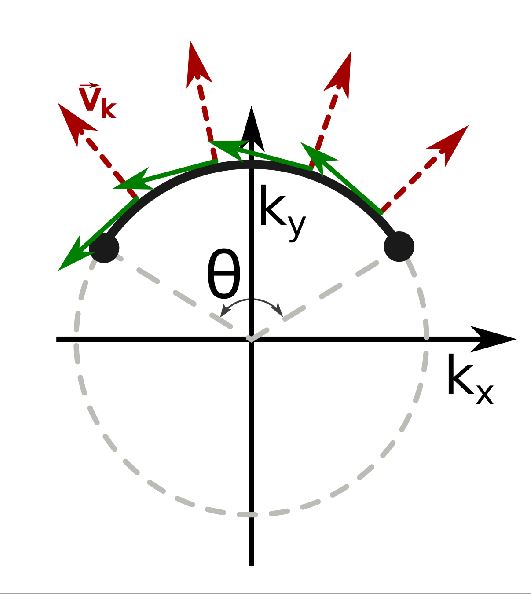}
\caption{ Model for computing the transport coefficient $\protect\lambda %
_{tr}$ along a curved arc of length $s$. The path consists of a segment of a
circle of radius $r=s/\protect\theta $. The velocity of the particle is
assumed to have a constant magnitude $v_{F}$ and to be orientated
perpendicular to the arc as illustrated (dotted red arrows). The spin of the
particle (solid green arrows) is assumed to be tangential to the arc. }
\label{FigA1}
\end{figure}

Here we consider contributions to electron--phonon coupling arising from
scattering within a minimal model for a single Fermi arc illustrated in
Figure~\ref{FigA1}. The model consists of a segment of a circle of radius $%
k_{arc}=s_{arc}/\theta _{arc}$ which subtends an angle $\theta _{arc}$ and
ends at a pair of the Weyl points. For our calculations, the length of the
arc $s_{arc}$ is assumed to be constant. Hence, the angle $\theta _{arc}$ is
related to the curvature of the arc. The particle is assumed to have the
relativistic dispersion relation $\epsilon _{\mathbf{k}}^{arc}=v_{F}k_{arc}$
along the arc. Hence, the velocity of the particle has a constant magnitude $%
v_{\mathbf{k}}=v_{F}$ and is orientated perpendicular to the arc. For the
Weyl point, we assume they are located at the ends of the arc with each
having the dispersion $\epsilon _{\mathbf{k}}^{WP}=v_{F}|\mathbf{k}-\mathbf{k%
}_{WP}|.$

\begin{widetext}

Consider electric field along the $y$ axis which is the axis of symmetry for
the arc. Recall that within the Boltzman theory, the electron--phonon
resistivity%
\begin{equation*}
\rho (T)=\frac{4\pi T}{\langle N\left( \epsilon _{F})v_{F\alpha }^{2}\right)
\rangle }\int_{0}^{\infty }\frac{d\omega }{\omega }\alpha _{tr}^{2}F(\omega
)\left( \frac{\omega /2T}{\sinh (\omega /2T)}\right) ^{2}
\end{equation*}%
is expressed via the so called transport spectral function $\alpha
_{tr}^{2}F(\omega )$ as follows (we use atomic units with $e=\hbar =m_{e}=1)$
\begin{equation*}
\alpha _{tr}^{2}F(\omega )=\frac{1}{\langle N\left( 0)v_{Fy}^{2}\right)
\rangle }\sum_{\nu }\sum_{\mathbf{k}j\mathbf{k}^{\prime }j^{\prime }}(v_{y%
\mathbf{k}j}-v_{y\mathbf{k}^{\prime }j^{\prime }})^{2}|V_{\mathbf{k}j\mathbf{%
k}^{\prime }j^{\prime }}^{e-ph}|^{2}\delta (\epsilon _{\mathbf{k}j})\delta
(\epsilon _{\mathbf{k\prime }j^{\prime }})\delta (\omega -\omega _{\mathbf{%
k-k}^{\prime }\nu })
\end{equation*}%
where $\omega _{\mathbf{k-k}^{\prime }\nu }$ are the phonon frequencies, $%
|V_{\mathbf{k}j\mathbf{k}^{\prime }j^{\prime }}^{e-ph}|^{2}$ are the
absolute squares of the electron--phonon matrix elements and 
\begin{equation*}
\langle N\left( 0)v_{Fy}^{2}\right) \rangle =\sum_{\mathbf{k}j}v_{y\mathbf{k}%
j}^{2}\delta (\epsilon _{\mathbf{k}j})
\end{equation*}%
are the mean squares of the Fermi velocities. In the following assume the
Debye approximation with $\omega _{\mathbf{q}\nu }=v_{s}q$ for $0<q<q_{\max
}\ $with the maximum wavevector $q_{\max }=\omega _{D}/v_{s}$ determined by
the Debye frequency $\omega _{D}$ and the sound velocity $v_{s}.$ Also
assume that the matrix element of the electron--phonon scattering depends
only on the transferred momentum%
\begin{equation*}
V_{\mathbf{k}j\mathbf{k}^{\prime }j^{\prime }}^{e-ph}=V_{jj^{\prime
}}^{e-ph}(\mathbf{k}-\mathbf{k}^{\prime })
\end{equation*}%
as this is the case for the deformation potential type model where 
\begin{equation*}
V_{\mathbf{k}j\mathbf{k}^{\prime }j^{\prime }}^{e-ph}=V^{DP}(\mathbf{k}-%
\mathbf{k}^{\prime })=D(\mathbf{k}-\mathbf{k}^{\prime })^{1/2}
\end{equation*}%
or including the Debye screening 
\begin{equation*}
V_{\mathbf{k}j\mathbf{k}^{\prime }j^{\prime }}^{e-ph}=\frac{D(\mathbf{k}-%
\mathbf{k}^{\prime })^{1/2}}{|\mathbf{k}-\mathbf{k}^{\prime }|^{n}+\kappa
_{D}^{n}}
\end{equation*}%
where $n=1$ or $2$ for 2D/3D cases respectively. With these approximations
we obtain%
\begin{equation*}
\alpha _{tr}^{2}F(\omega )=\frac{3\theta (\omega _{D}-\omega )}{\langle
N\left( \epsilon _{F})v_{Fy}^{2}\right) \rangle }\sum_{\mathbf{q}}\delta
(\omega -v_{s}q)|V^{e-ph}(\mathbf{q})|^{2}P_{tr}(\mathbf{q})
\end{equation*}%
where the transport phase space $P_{tr}(\mathbf{q})$ is the integral over
the Brillouin Zone%
\begin{equation*}
P_{tr}(\mathbf{q})=\sum_{\mathbf{k}}(v_{y\mathbf{k}}-v_{y\mathbf{k}+\mathbf{q%
}})^{2}\delta (\epsilon _{\mathbf{k}})\delta (\epsilon _{\mathbf{k}+\mathbf{q%
}})
\end{equation*}%
We can include spin dependence of the matrix elements assuming, for example,
that the spins are tangential to the arc. By representing 
\begin{equation*}
V^{e-ph}(q)=V^{Space}(q)V^{spin}(q)
\end{equation*}%
we assume that 
\begin{equation*}
V^{spin}(q)=\cos (\frac{\pi }{2}\frac{q}{2k_{arc}})
\end{equation*}%
Although this may not be the case for the straight arc with isotropic bulk
Weyl points of opposite chirality, such alignment will result in a correct
limit when the arc turns into the full circle with the spin momentum locking
characteristic of a 3D TI.

Consider now three cases of intra--arc scattering, scattering within the
Weyl points and arc--to-bulk scattering.

\subsection{Intra--arc scattering.}

All integrals are taken in 2D. The area unit cell is given by $A_{c}.$ We
first obtain%
\begin{eqnarray*}
N\left( 0\right) &=&\sum_{\mathbf{k}}\delta (\epsilon _{\mathbf{k}})=\frac{%
A_{c}}{(2\pi )^{2}}\frac{k_{arc}\theta _{arc}}{v_{F}}=\frac{A_{c}}{(2\pi
)^{2}}\frac{s_{arc}}{v_{F}} \\
\langle N(0)v_{y}\rangle &=&\sum_{\mathbf{k}}v_{\mathbf{k}y}\delta (\epsilon
_{\mathbf{k}})=\frac{A_{c}}{(2\pi )^{2}}2k_{arc}\sin \frac{\theta _{arc}}{2}=%
\frac{A_{c}}{(2\pi )^{2}}s_{arc}j_{0}(\frac{\theta _{arc}}{2}) \\
\langle N(0)v_{y}^{2}\rangle &=&\sum_{\mathbf{k}}v_{\mathbf{k}y}^{2}\delta
(\epsilon _{\mathbf{k}})=\frac{A_{c}}{(2\pi )^{2}}v_{F}k_{arc}\left[ \frac{1%
}{2}\theta _{arc}+\frac{1}{2}\sin \theta _{arc}\right] =\frac{A_{c}}{(2\pi
)^{2}}v_{F}s_{arc}\frac{1}{2}\left[ 1+j_{0}(\theta _{arc})\right]
\end{eqnarray*}

For the transport phase space integral we obtain in polar coordinates where $%
\phi _{q}$ is the angle with respect to $x$%
\begin{eqnarray*}
P^{tr}\left( q,\phi _{q}\right) &=&\frac{A_{c}}{(2\pi )^{2}}\frac{q\sin
^{2}\phi _{q}}{k_{arc}}\frac{1}{\sqrt{1-\left( \frac{q}{2k_{arc}}\right) ^{2}%
}}\theta (2k_{arc}\sin \theta _{arc}/2-q) \\
&&[\theta (\phi _{\max }(q)-\phi _{q})\theta (\phi _{q}+\phi _{\max
}(q))+\theta (\phi _{\max }(q)+\pi -\phi _{q})\theta (\phi _{q}+\phi _{\max
}(q)-\pi )]
\end{eqnarray*}%
where the maximum scattering angle is given by%
\begin{equation*}
\phi _{\max }(q)=\frac{\theta _{arc}}{2}-\arcsin \frac{q}{2k_{arc}}
\end{equation*}%
The expression is valid for $\theta _{arc}<\pi .$ For the transport spectral
function we obtain%
\begin{eqnarray*}
\alpha _{tr}^{2}F(\omega ) &=&\frac{A_{c}}{(2\pi )^{2}}\frac{1}{\frac{1}{2}%
v_{F}s_{arc}\left[ 1+j_{0}(\theta _{arc})\right] }\frac{\theta _{arc}}{%
v_{s}^{4}s_{arc}}\frac{6\omega ^{3}}{\sqrt{1-\left( \frac{\omega \theta
_{arc}}{2s_{arc}v_{s}}\right) ^{2}}}\left[ \frac{D^{2}\cos ^{2}(\frac{\pi }{2%
}\frac{\omega \theta _{arc}}{2s_{arc}v_{s}})}{(\omega \theta
_{arc}/2s_{arc}v_{s})+\kappa _{D}}\right] \\
&&\times \left[ \frac{\theta _{arc}}{2}-\arcsin \frac{\omega \theta _{arc}}{%
2s_{arc}v_{s}}-\frac{1}{2}\sin (\theta _{arc}-2\arcsin \frac{\omega \theta
_{arc}}{2s_{arc}v_{s}})\right] \theta (\omega _{D}-\omega )\theta
(s_{arc}v_{s}j_{0}(\theta _{arc}/2)-\omega )
\end{eqnarray*}%
where the expression in square brackets assumes possible forms of the
electron phonon matrix elements.

The resistivity behavior can be analyzed for the two regimes $T\gg T_{BG}$
and $T\ll T_{BG}$ set by the Bloch Gr\"{u}neisen temperature ($\theta
_{arc}<\pi )$ 
\begin{equation*}
T_{BG}=\min (\omega _{D},s_{arc}v_{s}j_{0}(\theta _{arc}/2))
\end{equation*}%
For the regime $T\gg T_{BG},$ we obtain the resistivity linear in $T$%
\begin{equation*}
\rho (T\gg T_{BG})\sim \lambda _{tr}T
\end{equation*}%
with the transport constant 
\begin{equation*}
\lambda _{tr}=\int_{0}^{\infty }d\omega \alpha _{tr}^{2}F(\omega )/\omega
\end{equation*}%
Note that $\lambda _{tr}(\theta _{arc})\sim \theta _{arc}^{2}$ and
disappears in the limit of the straight arc.

For the regime $T\ll T_{BG},$ the resistivity behavior is determined by the
low frequency expansion of $\alpha _{tr}^{2}F(\omega )\sim \omega ^{3}.$ We
obtain the resistivity 
\begin{equation*}
\rho (T\ll T_{BG})\sim T^{4}
\end{equation*}

The generalization for the angles $\pi <\theta _{arc}<2\pi $ is
straightforward. In particular, we give the answers for the case of the
circle with the radius $k_{F}.$ We obtain%
\begin{equation*}
P^{tr}\left( q,\phi _{q}\right) =\frac{A_{c}}{(2\pi )^{2}}\frac{2q}{k_{F}}%
\frac{\cos ^{2}\phi _{q}}{\sqrt{1-\left( \frac{q}{2k_{F}}\right) ^{2}}}%
\theta (2k_{F}-q)
\end{equation*}

\begin{equation*}
\alpha _{tr}^{2}F(\omega )=\frac{A_{c}}{(2\pi )^{2}}\frac{2}{%
k_{F}^{2}v_{F}v_{s}^{4}}\frac{\omega ^{3}}{\sqrt{1-\left( \frac{\omega }{%
2k_{F}v_{s}}\right) ^{2}}}\left[ \frac{D^{2}\cos ^{2}(\frac{\pi }{2}\frac{%
\omega \theta _{arc}}{2s_{arc}v_{s}})}{(\omega \theta
_{arc}/2s_{arc}v_{s})+\kappa _{D}}\right] \theta (\omega _{D}-\omega )\theta
(2k_{F}v_{s}-\omega )
\end{equation*}

Here the Bloch--Gr\"{u}neisen temperature is set by

\begin{equation*}
T_{BG}=\min (\omega _{D},2k_{F}v_{s})
\end{equation*}

In the limit $T\gg T_{BG}$ we obtain the resistivity linear in $T$%
\begin{equation*}
\rho (T\gg T_{BG})\sim \lambda _{tr}T
\end{equation*}%
Note here that the inclusion of the spin dependence set by $V_{spin}(q)$ in
the matrix element reduces the coupling constant $\lambda _{tr}$ by 2--5
times depending on a particular form of the spatial matrix element.

For the regime $T\ll T_{BG},$ the resistivity behavior is determined by the
low frequency expansion of $\alpha _{tr}^{2}F(\omega )\sim \omega ^{3}.$ We
again obtain the resistivity $\rho (T\ll T_{BG})\sim T^{4}.$

\subsection{Weyl Point Scattering}

All quantities are now referred to the unit cell $\Omega _{c}$ of the 3D
solid. For the bulk scattering within a single Weyl point set by the
isotropic dispersion $\epsilon _{\mathbf{k}}^{WP}=v_{F}k$ in 3D we obtain

\begin{eqnarray*}
N\left( 0\right) &=&\frac{\Omega _{c}}{(2\pi )^{3}}\frac{4\pi k_{F}^{2}}{%
v_{F}} \\
\langle N(0)v_{\alpha }\rangle &=&0 \\
\langle N(0)v_{\alpha }^{2}\rangle &=&\frac{\Omega _{c}}{(2\pi )^{3}}\frac{%
4\pi }{3}v_{F}k_{F}^{2}
\end{eqnarray*}%
Here the Fermi wave vector $k_{F}$ assumes a doping away from the nodal
point. The transport phase space integral in spherical coordinates with $%
\theta _{q}$ measured away from z axis we obtain 
\begin{equation*}
P^{tr}\left( q,\theta _{q}\right) =\frac{\Omega _{c}}{(2\pi )^{3}}2\pi q\cos
^{2}\theta _{q}\theta \left( 2k_{F}-q\right)
\end{equation*}%
For the transport spectral function $\alpha _{tr}^{2}F(\omega )$ assume for
simplicity the deformation potential electron--phonon matrix element $%
|V^{e-ph}(q)|^{2}=D^{2}q.$ The result is given by 
\begin{equation*}
\alpha _{tr}^{2}F(\omega )=\frac{\Omega _{c}}{(2\pi )^{3}}\frac{6\pi
D^{2}\omega ^{4}}{v_{s}v_{F}k_{F}^{2}}\theta (2k_{F}v_{s}-\omega )\theta
(\omega _{D}-\omega )
\end{equation*}%
The Bloch--Gr\"{u}neisen temperature is set by

\begin{equation*}
T_{BG}=\min (\omega _{D},2k_{F}v_{s})
\end{equation*}

In the limit $T\gg T_{BG}=\min (\omega _{D},2k_{F}v_{s})$ we obtain for the
bulk resistivity per volume%
\begin{equation*}
\rho \left( T\right) =\frac{6\pi T}{4}\frac{D^{2}}{v_{s}v_{F}^{2}k_{F}^{4}}%
[\min (\omega _{D},2k_{F}v_{s})]^{4}
\end{equation*}%
It shows the behavior linear in $T.$

Note that if the $2k_{F}<\omega _{D}/v_{s}$ (in the vicinity of the nodal
point) we obtain $k_{F}$ independent behavior 
\begin{equation*}
\rho (T)=\frac{D^{2}v_{s}^{3}}{v_{F}^{2}}24\pi T
\end{equation*}

In the limit $T\ll \min (\omega _{D},2k_{F}v_{s})$, the $\sinh $ is very
small for $\omega \gg \min (\omega _{D},2k_{F}v_{s})$ and we can set the
upper limit of integration to $2T.$ The resistivity shows $T^{5}$ behavior%
\begin{equation*}
\rho _{\alpha }(T)=6\pi T\frac{D^{2}(2T)^{4}}{v_{s}v_{F}^{2}k_{F}^{4}}%
\int_{0}^{1}\frac{x^{5}dx}{\sinh ^{2}(x)}
\end{equation*}%
where the value of the integral here is 0.202. Note the resistivity grows as 
$k_{F}$ approaches the nodal point.

\subsection{Arc--To--Bulk Scattering.}

Here we have to either assume that we measure the conductivity within a
finite slab of the width $N_{z}a_{z}$ or we define the current within a thin
surface layer $N_{z}a_{z}$ by applying electric field only within this
layer, since in the bulk thermodynamic limit ($N_{z}\rightarrow \infty )$
the bulk conductivity will scale proportional to $N_{z}$ and overwhelm all
surface effects. The transport spectral function 
\begin{eqnarray*}
\alpha _{tr}^{2}F(\omega ) &=&\frac{1}{\langle N\left( \epsilon
_{F})v_{Fy}^{2}\right) \rangle }\sum_{\nu }\sum_{\mathbf{k}j\mathbf{k}%
^{\prime }j^{\prime }}(v_{y\mathbf{k}j}-v_{y\mathbf{k}^{\prime }j^{\prime
}})^{2}|V_{\mathbf{k}j\mathbf{k}^{\prime }j^{\prime }}^{e-ph}|^{2}\delta
(\epsilon _{\mathbf{k}j})\delta (\epsilon _{\mathbf{k\prime }j^{\prime
}})\delta (\omega -\omega _{\mathbf{k-k}^{\prime }\nu }) \\
&=&[\alpha _{tr}^{2}F(\omega )]_{WP}+[\alpha _{tr}^{2}F(\omega
)]_{arc}+[\alpha _{tr}^{2}F(\omega )]_{arc\longleftrightarrow WP}
\end{eqnarray*}%
includes transitions $\mathbf{k}j\rightarrow \mathbf{k}^{\prime }j^{\prime }$
within the Weyl points projected onto the slab Brillouin Zone, $[\alpha
_{tr}^{2}F(\omega )]_{WP},$ within the arc, $[\alpha _{tr}^{2}F(\omega
)]_{WP},$ and the arc--to--WP transitions, $[\alpha _{tr}^{2}F(\omega
)]_{arc\longleftrightarrow WP}$. Note that the normalization factor here is
proportional to the slab size 
\begin{equation*}
\langle N\left( 0)v_{Fy}^{2}\right) \rangle =\frac{N_{z}\Omega _{c}}{(2\pi
)^{3}}\frac{4\pi }{3}v_{F}k_{F}^{2}+\frac{A_{c}}{(2\pi )^{2}}v_{F}s_{arc}%
\frac{1}{2}\left[ 1+j_{0}(\theta _{arc})\right]
\end{equation*}%
where we approximate the contribution from the Weyl point by its bulk value
per unit cell times the number of unit cells in the slab given by $N_{z.}$
We note that $[\alpha _{tr}^{2}F(\omega )]_{WP}$ includes the transitions $%
j\rightarrow j^{\prime }$ between various projections of the same bulk state 
$|\mathbf{k}\rangle _{WP}$ now appeared in the surface Brillouin Zone, but
they are orthogonal to each other. This will collapse the double sum to a
single sum over $j$ which is proportional to $N_{z}$. At the absence of the
surface scattering, such scaling in the numerator will be cancelled with$%
~N_{z}$ appeared in the denominator and $[\alpha _{tr}^{2}F(\omega )]_{WP}$
will not depend on $N_{z}.$ The resistivity has additional prefactor of $%
1/\langle N\left( 0)v_{Fy}^{2}\right) \rangle $ therefore will scale as $%
1/N_{z}$ which simply gives the result that the conductivity of the slab is
the conductivity per unit cell volume times the number of unit cells in the
slab. For the finite $N_{z},$ the transitions $|\mathbf{k}j\rangle
_{WP}\rightarrow |\mathbf{k}^{\prime }j\rangle _{WP}$ within the Weyl point
will disappear when $k_{F}\rightarrow 0$ and so will the arc--to--WP
transitions. Therefore the resistivity will be determined by the intra--arc
scattering alone. This result will hold as long as $N_{z}$ is finite.

\end{widetext}

\section{Appendix B. Lattice Models for Weyl Semimetal and Topological
Insulator}

We consider the following minimal two--band 3D Weyl semimetal model on a
simple cubic lattice \cite{Turner}, 
\begin{align}
\hat{H}_{Weyl}(\mathbf{k})& =2t_{x}(\cos (k_{x}a_{x})-\cos (k_{0}a_{x})) 
\notag \\
& +m(2-\cos (k_{y}a_{y})-\cos (k_{z}a_{z}))\hat{\sigma}_{x}  \notag \\
& +2t_{y}\sin (k_{y}a_{y})\hat{\sigma}_{y}+2t_{z}\sin (k_{z}a_{z})\hat{\sigma%
}_{z},  \label{eq:wsm_h_bulk}
\end{align}%
The model breaks time reverse symmetry and contains two Weyl nodes of
opposite chirality at $\mathbf{k}=(\pm k_{W},\,0,\,0)$. For all of our
calculations we again set the lattice constants to unity, $a_{i}=1,$ and use
the parameters $m=0.1$, $t_{x}=-0.05$, $t_{y}=t_{z}=0.05$, and $k_{W}=\pi /4$%
. To simulate the Fermi arc surface states we use a cubic lattice which is
infinite along $xy-$axis but which has $N=41$ layers along the z--axis. The
Fermi arc is the straight line which connects the two Weyl nodes at $%
k_{W}=\pm \pi /4$ in momentum space.

\begin{figure}[tbp]
\includegraphics[width=0.3\textwidth]{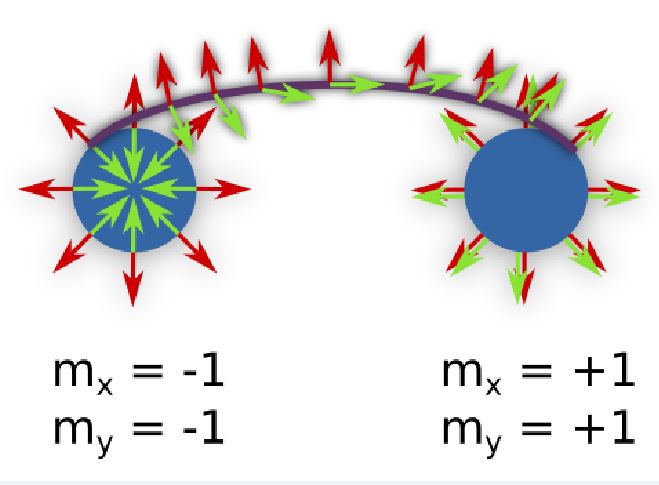} \includegraphics[width=0.3%
\textwidth]{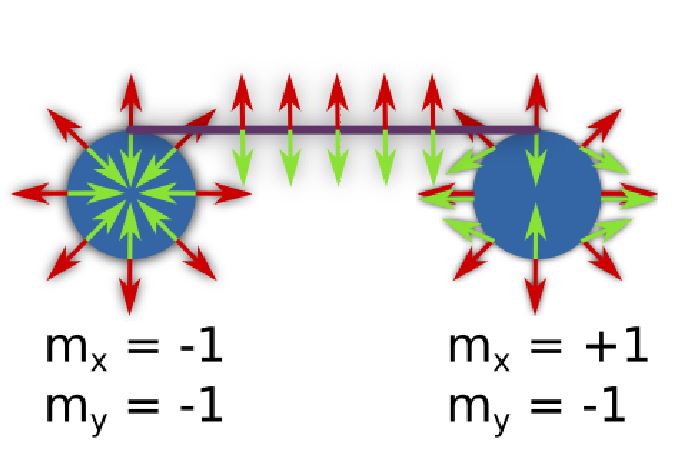}
\caption{The velocity (red arrows) and spin (green arrows) distribution for
the surface states of two different WSM states which have a Fermi energy
slightly above the Weyl points. In the case of WSM (a), the Weyl points
(blue circles) have different values for $m_{x}$ and $m_{y}$ in the Weyl
Hamiltonian, Eq.\protect\ref{HWeyl}. Since the Fermi arc (purple lines) must
smoothly merge with the Weyl points, the spin orientation of the Fermi arc
near each Weyl point must approach the spin orientation at the Weyl point.
In the case of WSM (b), the Weyl points have different values for $m_{x}$
and the same values for $m_{y}$. In the limit of a straight Fermi arc, the
states along the entire length of the arc can have the same spin
orientation. }
\label{FigB1}
\end{figure}

In the vicinity of a Weyl point the Hamiltonian can be written in a general
form\cite{Model} 
\begin{equation}
\hat{H}_{Weyl}=\sum_{\alpha \in \{x,y,z\}}m_{\alpha }|v_{\alpha }|k_{\alpha
}\sigma _{\alpha },  \label{HWeyl}
\end{equation}%
where $m_{\alpha }$ is either positive or negative unity, $v_{\alpha }$ is
the velocity of the particle along the respective axis, and the chirality of
the Weyl node is the product $m_{x}m_{y}m_{z}.$ Usually, isotropic case is
discussed, where $m_{x}=m_{y}=m_{z}$ the electron spin along the arc must
change its orientation (at least once) while traversing between the Weyl
nodes of opposite chirality ( see Figure \ref{FigB1}a). \ However, our
minimal model assumes a more general anisotropic case where spin retains the
same orientation along the entire length of the arc (see Figure \ref{FigB1}%
b).

The following minimal four--band model is used to simulate a 3D Topological
insulator on a simple cubic lattice with nearest--neighbor hopping~\cite%
{TIShen}, 
\begin{align}
\hat{H}_{TI}(\mathbf{k})& =A\left[ \sum_{i=x,y,z}\sin (k_{i}a_{i})\hat{\alpha%
}_{i}\right]  \notag \\
& +\left[ \Delta -4B\left[ \sum_{i=x,y,z}\sin ^{2}\left( \frac{k_{i}a_{i}}{2}%
\right) \right] \right] \hat{\beta},  \label{eq:ti_h_bulk}
\end{align}%
where the Dirac matrices are given in terms of the Pauli matrix via the
relations, 
\begin{equation*}
\hat{\alpha}_{i}=\hat{\sigma}_{x}\otimes \hat{\sigma}_{i},\hat{\beta}=\hat{%
\sigma}_{z}\otimes \hat{\sigma}_{0}.
\end{equation*}%
For all of our simulations we set the lattice constants to unity, $a_{i}=1$
and use the parameters $A=B=\Delta =0.1$. To simulate surface states, we use
a cubic lattice which is infinite along the $xy$--axis but which has $N=40$
sites along the z--axis with two spins per each site. The band structure
contains a doubly degenerate 2D Dirac cone centered at the $\Gamma $ point
consisting of states that are exponentially localized on opposing surfaces
of the slab structure. The bulk band structure is fully gapped.

These parameters were chosen to insure that the velocity of the TI Dirac
cone and the velocity of the WSM Fermi arc are approximately equivalent. In
the case of TI, the Fermi level matches the density of states of WSM\ at $%
\epsilon _{F}=0.02.$ We explored two levels close to the Dirac point ($%
\epsilon _{F}=0.02$ and $\epsilon _{F}=0.05$) while in the case of WSM the
Fermi level was taken to be pinned to the Weyl nodes ($\epsilon _{F}=0.0$).

For the simulations with disorder a complex energy--dependent self--energy $%
\Sigma (\omega )$ is obtained by using a Coherent Potential Approximation
which is then utilized in calculating the conductivities using
Kubo--Greenwood formalism. For integrals over the Brillouin Zone, that
appear in Kubo--Greenwood calculation, we use two complementary methods, a
simple k--point summation and a tetrahedron method. For simple k--sums, we
found that we need to use enormous numbers of grid points, up to 4000x4000,
to reach complete convergency of the calculated conductivities. We also
developed a new tetrahedron method for integrating products of two Green
functions exhibiting two poles in a complex plane. The method has been
previously used in its simplified version\cite{DMFTtetra, KH} to calculate
transport and optical properties for strongly correlated systems, and is now
generalized for arbitrary cases of energy band degeneracies at the corners
of the tetrahedron. We found that the tetrahedron method allows us to use
much coarser grids, 120x120, in order to reproduce the results obtained from
the fully convergent k--sum integration.

\section{Appendix C. Simulation of surface vacancies for TaAs using Coherent
Potential Approximation}

We determine the one--electron energy states of 6 unit--cell (24 atomic
layers) slab structure using density functional theory (DFT) with
generalized gradient approximation (GGA) for the exchange--correlation
potential\cite{GGA} as implemented in the full potential linear muffin--tin
orbital (FP--LMTO) method\cite{Savrasov}. In order to perform simulation of
vacancies on the topmost As layer, we implement a coherent potential
approximation, a self--consistent dynamical mean field approach that allows
to extract disorder induced self--energies $\hat{\Sigma}_{CPA}(\omega )$
from the FP--LMTO\ calculation

The Coherent Potential Approximation is a self--consistent method that
allows to determine $\hat{\Sigma}_{CPA}(\omega )$ by reducing the problem to
a single scatterer problem embedded into the effective medium. It relies on
separating the single particle Hamiltonian onto the hopping term $T$ and
on--site scattering term $V$. For two pure systems A and B this implies the
forms%
\begin{equation*}
H^{\mathbf{k}}(A)=T^{\mathbf{k}}+V(A)
\end{equation*}%
\begin{equation*}
H^{\mathbf{k}}(B)=T^{\mathbf{k}}+V(B)
\end{equation*}%
With a given initial guess for self--energy $\hat{\Sigma}_{CPA}(\omega ),$
the local Green function of the disordered medium $A_{1-x}B_{x}$ is computed 
\begin{equation*}
\hat{G}_{loc}(\omega )=\sum_{\mathbf{k}}\left( \omega \hat{I}-\hat{T}^{%
\mathbf{k}}-\hat{\Sigma}_{CPA}(\omega )\right) ^{-1}
\end{equation*}%
It defines the so called bath Green's function 
\begin{equation*}
\mathcal{G}^{-1}(\omega )=\hat{G}_{loc}^{-1}(\omega )+\hat{\Sigma}%
_{CPA}(\omega )
\end{equation*}%
containing hybridization effects of the local scatterer with the medium. The
scatterers A and B embedded into the medium are now described by the
"impurity" Green functions%
\begin{equation*}
G_{A/B}(\omega )=\left[ \mathcal{G}^{-1}(\omega )-V_{A/B}\right] ^{-1}
\end{equation*}%
from which the effective scatterer of the disordered medium is simply viewed
as the weighted average%
\begin{equation*}
(1-x)G_{A}(\omega )+xG_{B}(\omega )=\left[ \mathcal{G}^{-1}(\omega )-\hat{%
\Sigma}_{CPA}(\omega )\right] ^{-1}
\end{equation*}%
This equation defines new self--energy $\hat{\Sigma}_{CPA}(\omega )$ at the
iteration that is then used to compute new $\hat{G}_{loc}(\omega )$ and the
entire procedure is repeated until self--consistency over $\hat{\Sigma}%
_{CPA}(\omega )$ (or equivalently over $\mathcal{G}(\omega )$) is reached.

Our implementation of these CPA\ equations is based on their combination
with Density Functional Theory using a projector operator technique similar
to as strongly correlated materials are studied using a combination of DFT
and dynamical mean field theory (DMFT) methods\cite{DMFT}. To set up the
DFT+CPA\ method, we define a site $\tau $ dependent projector operator 
\begin{equation*}
\sum_{\alpha \beta }|\phi _{\alpha \tau }\rangle \langle \phi _{\beta \tau }|
\end{equation*}%
with help of radial solutions $|\phi _{\alpha }\rangle $ of the
one--electron Schroedinger equation taken with a spherically symmetric part
of the full potential. The disorder induced dynamical self--energy at site $%
\tau $ is now defined with help of the projectors%
\begin{equation*}
\hat{\Sigma}_{CPA}(\omega )=\sum_{\alpha \beta }|\phi _{\alpha \tau }\rangle
\Sigma _{\alpha \beta }^{\tau }(\omega )\langle \phi _{\beta \tau }|
\end{equation*}%
and used to construct local Green's function of the lattice%
\begin{equation*}
G_{\alpha \beta }^{\tau }(\omega )=\sum_{\mathbf{k}}\left[ \langle \chi ^{%
\mathbf{k}}|\omega \hat{I}-\hat{H}-\hat{\Sigma}_{CPA}(\omega )|\chi ^{%
\mathbf{k}}\rangle \right] _{\alpha \beta }^{-1}
\end{equation*}%
where the matrix elements of the $\hat{H}+\hat{\Sigma}_{CPA}$ are taken
using the LMTO\ basis set $|\chi _{\alpha }^{\mathbf{k}}\rangle .$

We now define scattering potentials $V_{A}$ and $V_{B}$ in reference to pure
systems. For the problem at hand, we treat the original slab structure of
TaAs as a pure system A, and the same slab structure with top As layer
removed as a pure system B. This can be achieved by adding a very large
potential to each surface As atom that effectively un--hybridizes the top As
layer from the rest of the slab. The vacancy induced self--energy $\hat{%
\Sigma}_{CPA}$ is viewed as the correction to the Hamiltonian of pure system
A, hence we set $V_{A}=0,$ and $V_{B}=\infty $ (technically, a very large
number, 10,000). The self--consistent CPA procedure is then utilized after
which the surface Greens functions (slab Green functions projected onto 4
topmost atomic layers) are visualized on Fig. 8 of the main text.


\begin{thebibliography}{99}
\bibitem{Volovik} For a review, see, e.g., G. E. Volovik, The Universe in a
Helium Droplet, (Clarendon, Oxford, England, 2003).

\bibitem{Nielsen} H. B. Nielsen and M. Ninomiya, Phys. Lett. 130 B, 389
(1983).

\bibitem{Wan} X. Wan, A. M. Turner, A. Vishwanath, and S. Y. Savrasov, Phys.
Rev. B 83, 205101 (2011).

\bibitem{Yang} Kai-Yu Yang, Yuan-Ming Lu, and Ying Ran, Physical Review B
84, 075129 (2011).

\bibitem{Burkov} A. A. Burkov, L. Balents, Phys. Rev. Lett. 107, 127205
(2011).

\bibitem{XiDaiTaAs} B. Q. Lv, H. M. Weng, B. B. Fu, X.P. Wang, H. Miao, J.
Ma, P. Richard, X. C. Huang, L. X. Zhao, G. F. Chen, Z. Fang, X. Dai, T.
Qian, and H. Ding, Phys. Rev. X 5, 031013 (2015).

\bibitem{HasanTaAs} Su-Yang Xu, Ilya Belopolski, Nasser Alidoust, Madhab
Neupane, Guang Bian, Chenglong Zhang, Raman Sankar, Guoqing Chang, Zhujun
Yuan, Chi-Cheng Lee, Shin-Ming Huang, Hao Zheng, Jie Ma, Daniel S Sanchez,
BaoKai Wang, Arun Bansil, Fangcheng Chou, Pavel P Shibayev, Hsin Lin, Shuang
Jia, M Zahid Hasan, Science 349, 613 (2015).

\bibitem{HosurReview} For a review, see, e.g., Pavan Hosur, Xiaoliang Qi,
Comptes Rendus Physique, 14, 857 (2013).

\bibitem{AshvinReview} For a review, see, e.g., N.P. Armitage, E. J. Mele,
Ashvin Vishwanath (unpublished).

\bibitem{HasanRMP} For a review, see, e.g., M. Z. Hasan and C. L. Kane, Rev.
Mod. Phys. 82, 3045 (2010).

\bibitem{MooreNature} For a review, see, e.g., J. E. Moore, Nature 464, 194
(2010).

\bibitem{PhysicsToday} For a review, see, e.g., Xiao-Liang Qi and Shou-Cheng
Zhang, Physics Today 63, 33 (2010).

\bibitem{NNTheorem} Nielsen, H.B.; Ninomiya, M. Phys. Lett. B 105: 219
(1981).

\bibitem{DSM} S. A. Parameswaran, T. Grover, D. A. Abanin, D. A. Pesin, and
A. Vishwanath, Phys. Rev. X 4, 031035 (2014).

\bibitem{QuantumOscillations} Andrew C. Potter, Itamar Kimchi \& Ashvin
Vishwanath, Nature Comm. 5, 6161 (2014).

\bibitem{AshvinTransport} Pavan Hosur, S. A. Parameswaran, and Ashvin
Vishwanath, Charge Transport inWeyl Semimetals, Phys. Rev. Lett. 108, 046602
(2012).

\bibitem{Balatsky} Zhoushen Huang, Tanmoy Das, Alexander V. Balatsky, and
Daniel P. Arovas, Phys. Rev. B 87, 155123 (2013).

\bibitem{DiffusiveTransport} Rudro R. Biswas, and Shinsei Ryu, Physical
Review B 89, 014205 (2014).

\bibitem{QuantumTransport} Bj\"{o}rn Sbierski, Gregor Pohl, Emil J.
Bergholtz, and Piet W. Brouwer, Phys. Rev. Lett. 113, 026602 (2014).

\bibitem{ALTransport} J. H. Pixley, Pallab Goswami, and S. Das Sarma, Phys.
Rev. Lett. 115, 076601 (2015).

\bibitem{CriticalTransport} S. V. Syzranov, L. Radzihovsky, and V. Gurarie, 
Phys. Rev. Lett. 114, 166601 (2015).

\bibitem{Carbotte} C. J. Tabert, J. P. Carbotte, and E. J. Nicol, Phys. Rev.
B 93, 085426 (2016).

\bibitem{Nandkishore} Nandkishore, R., D. A. Huse, and S. L. Sondhi, Phys.
Rev. B 89, 245110 (2014).

\bibitem{Pixley} Pixley, J. H., D. A. Huse, and S. Das Sarma, Phys. Rev. X
6, 021042 (2016).

\bibitem{ArcTransport} E. V. Gorbar, V. A. Miransky, I. A. Shovkovy and P.
O. Sukhachov,  Phys. Rev. B 93, 235127 (2016).

\bibitem{Kubo} See, e.g., G. D. Mahan, Many--Particle Physics Plenum, New
York (1990).

\bibitem{DasSarma} Dimitrie Culcer, E. H. Hwang, Tudor D. Stanescu, and S.
Das Sarma, Phys. Rev. B 82, 155457 (2010).

\bibitem{Ong} J. G. Checkelsky,Y. S. Hor, R. J. Cava, and N. P. Ong, Phys.
Rev. Lett. 106, 196801 (2011).

\bibitem{StrongDisorder} Gerald Schubert,Holger Fehske, Lars Fritz, and
Matthias Vojta, Phys. Rev. B 85, 201105 (2012).

\bibitem{Bi2Se3} Dohun Kim, Sungjae Cho, Nicholas P. Butch, Paul Syers,
Kevin Kirshenbaum, Shaffique Adam, Johnpierre Paglione and Michael S.
Fuhrer, Nature Phys. 8, 459 (2012).

\bibitem{Sinova} Xin Liu and Jairo Sinova, Phys. Rev.  Lett. 111, 166801
(2013).

\bibitem{Austin} Xingyue Peng, Yiming Yang, Rajiv R. P. Singh, Sergey Y.
Savrasov, Dong Yu, Nature Comm. 7, 10878 (2016).

\bibitem{CPAReview} For a review, see, e.g., Yonezawa, F. \& Morigaki, K.
Prog. Theor. Phys. Supp. 53, 1--76 (1973).

\bibitem{Efetov} D. K. Efetov, P. Kim, Phys. Rev. Lett. 105, 256805 (2010).

\bibitem{Fuhrer} M. S. Fuhrer, Physics 3, 106 (2010).

\bibitem{Allen} P. B. Allen, Phys. Rev. B 17, 3725 (1978).

\bibitem{SavrasovEPI} S. Y. Savrasov, D. Y. Svarasov and O. K. Andersen,
Phys. Rev. Lett. 72, 372 (1994).

\bibitem{Bi2Se3Transport} Yashina, L. V. et al. ACS Nano 7, 5181--5191
(2013).

\bibitem{Bi2Se3Transport2} Kong, D. S. et al. ACS Nano 5, 4698--4703 (2011).

\bibitem{AL} Kentaro Nomura, Mikito Koshino, and Shinsei Ryu, Phys. Rev.
Lett. 99, 146806 (2007).

\bibitem{Turner} A. M. Turner and A. Vishwanath, arXiv 1301.0330 (2013).

\bibitem{TIShen} S.-Q. Shen, Topological Insulators, vol. 174 of Springer
Series in Solid-State Sciences (Springer-Verlag Berlin Heidelberg, 2012).

\bibitem{RelativisticGas2D} E. H. Hwang and S. Das Sarma, Phys. Rev. B 75,
205418 (2007).

\bibitem{Chen} R. Y. Chen, S. J. Zhang, J. A. Schneeloch, C. Zhang, Q. Li,
G. D. Gu, and N. L. Wang, Phys. Rev. B 92, 075107 (2015).

\bibitem{S1} L. X. Yang, Z. K. Liu, Y. Sun, H. Peng, H. F. Yang, T. Zhang,
B. Zhou, Y. Zhang, Y. F. Guo, M. Rahn, D. Prabhakaran, Z. Hussain, S.-K. Mo,
C. Felser, B. Yan \& Y. L. Chen, Nature Phys. 11, 728 (2015).

\bibitem{S2} Chandra Shekhar, Ajaya K. Nayak, Yan Sun, Marcus Schmidt,
Michael Nicklas, Inge Leermakers, Uli Zeitler, Yurii Skourski, Jochen
Wosnitza, Zhongkai Liu, Yulin Chen, Walter Schnelle, Horst Borrmann, Yuri
Grin, Claudia Felser, Binghai Yan, Nature Phys. 11, 645 (2015).

\bibitem{S3} N. Xu, H. M. Weng, B. Q. Lv, C. E. Matt, J. Park, F. Bisti, V.
N. Strocov, D. Gawryluk, E. Pomjakushina, K. Conder, N. C. Plumb, M.
Radovic, G. Aut\`{e}s, O. V. Yazyev, Z. Fang, X. Dai, T. Qian, J. Mesot, H.
Ding, M. Shi, Nature Comm. 7, 11006 (2016).

\bibitem{S4} Su-Yang Xu, Nasser Alidoust, Ilya Belopolski, Zhujun Yuan,
Guang Bian, Tay-Rong Chang, Hao Zheng, Vladimir N. Strocov, Daniel S.
Sanchez, Guoqing Chang, Chenglong Zhang, Daixiang Mou, Yun Wu, Lunan Huang,
Chi-Cheng Lee, Shin-Ming Huang, BaoKai Wang, Arun Bansil, Horng-Tay Jeng,
Titus Neupert, Adam Kaminski, Hsin Lin, Shuang Jia, M. Zahid Hasan, Nature
Phys. 11, 748 (2015).

\bibitem{S5} Lunan Huang, Timothy M. McCormick, Masayuki Ochi, Zhiying Zhao,
Michi-To Suzuki, Ryotaro Arita, Yun Wu, Daixiang Mou, Huibo Cao, Jiaqiang
Yan, Nandini Trivedi, Adam Kaminski, Nature Materials 15, 1155 (2016).

\bibitem{S6} G. Aut\`{e}s, D. Gresch, M. Troyer, A.A. Soluyanov, and O.V.
Yazyev, Phys. Rev. Lett. 117, 066402 (2016).

\bibitem{XG1} Alexey A. Soluyanov, Dominik Gresch, Zhijun Wang, QuanSheng
Wu, Matthias Troyer, Xi Dai, B. Andrei Bernevig, Nature 527, 495 (2015).

\bibitem{S7} Su-Yang Xu, Nasser Alidoust, Guoqing Chang, Hong Lu, Bahadur
Singh, Ilya Belopolski, Daniel Sanchez, Xiao Zhang, Guang Bian, Hao Zheng,
Marius-Adrian Husanu, Yi Bian, Shin-Ming Huang, Chuang-Han Hsu, Tay-Rong
Chang, Horng-Tay Jeng, Arun Bansil, Vladimir N. Strocov, Hsin Lin, Shuang
Jia, M. Zahid Hasan,  arXiv:1603.07318 (2016).

\bibitem{S8} Xiaochun Huang, Lingxiao Zhao, Yujia Long, Peipei Wang, Dong
Chen, Zhanhai Yang, Hui Liang, Mianqi Xue, Hongming Weng, Zhong Fang, Xi
Dai, and Genfu Chen, Phys. Rev. X 5, 031023 (2015).

\bibitem{S9} Ilya Belopolski, Peng Yu, Daniel S. Sanchez, Yukiaki Ishida,
Tay-Rong Chang, Songtian S. Zhang, Su-Yang Xu, Daixiang Mou, Hao Zheng,
Guoqing Chang, Guang Bian, Horng-Tay Jeng, Takeshi Kondo, Adam Kaminski,
Hsin Lin, Zheng Liu, Shik Shin, M. Zahid Hasan, arXiv:1610.020013 (2016).

\bibitem{S10} J. Ruan, S.-K. Jian, H. Yao, H. Zhang, S.-C. Zhang, D. Xing,
Nature Comm. 7, 11136 (2016).

\bibitem{S11} Guoqing Chang, Su-Yang Xu3, Daniel S. Sanchez, Shin-Ming
Huang, Chi-Cheng Lee, Tay-Rong Chang, Guang Bian, Hao Zheng, Ilya
Belopolski, Nasser Alidoust, Horng-Tay Jeng, Arun Bansil, Hsin Lin, M. Zahid
Hasan, Science Advances 2, e1600295 (2016).

\bibitem{S12} Tay-Rong Chang, Su-Yang Xu, Guoqing Chang, Chi-Cheng Lee,
Shin-Ming Huang, BaoKai Wang, Guang Bian, Hao Zheng, Daniel S. Sanchez, Ilya
Belopolski, Nasser Alidoust, Madhab Neupane, Arun Bansil, Horng-Tay Jeng,
Hsin Lin, M. Zahid Hasan, Nature Comm. 7, 10639 (2016).

\bibitem{S13} Q. Xu, Z. Song, S. Nie, H. Weng, Z. Fang, and X. Dai, Physial.
Review B 92, 205310 (2015).

\bibitem{S14} Z. K. Liu, L. X. Yang, Y. Sun, T. Zhang, H. Peng, H. F. Yang,
C. Chen, Y. Zhang, Y. F. Guo, D. Prabhakaran, M. Schmidt, Z. Hussain, S.-K.
Mo, C. Felser, B. Yan, Y. L. Chen, Nature Materials 15, 27 (2016).

\bibitem{S15} Hiroyuki Inoue, Andr\'{a}s Gyenis, Zhijun Wang, Jian Li, Seong
Woo Oh, Shan Jiang, Ni Ni, B. Andrei Bernevig, Ali Yazdani, Science 351,
1184 (2016).

\bibitem{TwoTipsSTM} Q. Niu, M. C. Chang, and C. K. Shih, Phys. Rev. B 51,
5502 (1995).

\bibitem{Model} Pierre Delplace, Jian Li and David Carpentier, EPL 97, 67004
(2012).

\bibitem{DMFTtetra} V. S. Oudovenko, G. P\'{a}lsson, K. Haule, and G.
Kotliar, S. Y. Savrasov, Phys. Rev. B, 035120 (2006).

\bibitem{KH} Kristjan Haule, Chuck-Hou Yee, and Kyoo Kim, Phys. Rev. B 81,
195107 (2010).

\bibitem{GGA} J. P. Perdew, K. Burke, and M. Ernzerhof, Phys. Rev. Lett. 77,
3865 (1996).

\bibitem{Savrasov} S. Y. Savrasov, Phys. Rev. B 54, 16470 (1996).

\bibitem{DMFT} For a review, see, e.g., G. Kotliar, S. Y. Savrasov, K.
Haule, V. S. Oudovenko, O. Parcollet, C.A. Marianetti, Rev. Mod. Phys., 78,
865, (2006).
\end{thebibliography}
\end{document}